\documentstyle[aps,multicol,epsf]{revtex}
\newcommand{\be}{\begin{equation}}
\newcommand{\ee}{\end{equation}}
\begin{document}
\draft
\title{Stochastic Model for a Vortex Depinning in Random Media \\} 
\author{Byungnam Kahng, Kwangho Park and Jinhee Park \\}
\address{
Department of Physics and Center for Advanced Materials 
and Devices,\\
Kon-Kuk University, Seoul 143-701, Korea \\}

\maketitle
\thispagestyle{empty}

\begin{abstract}
We present a self-organized stochastic model for the 
dynamics of a single flux line in random media. 
The dynamics for the flux line in the longitudinal 
and the transversal direction to an averaged moving 
direction are coupled to each other. 
The roughness exponents of the flux line are measured 
for each direction, which are close to 
$\alpha_{\parallel}\approx 0.63$ for the longitudinal and 
$\alpha_{\perp}\approx 0.5$ for the transversal direction, 
respectively. The dynamic exponents are obtained 
as $z\approx 1$ for both directions. We discuss the 
classification of universality for the stochastic model. 
\end{abstract}


\begin{multicols}{2}
\narrowtext
In the past few years, there has been an explosion of studies in the 
field of dynamics of fluctuating interfaces due to theoretical interests 
in the classification of universality for stochastic models and also 
due to applications to various physical phenomena such as crystal growth, 
vapor deposition, electroplating, biological growth, $etc$. 
A number of discrete models and continuum equations for interface 
dynamics have been introduced and studied [1-3]. 
An interesting feature of nonequilibrium interface dynamics  
is the nontrivial dynamic scaling behavior~\cite{family} 
of the interface fluctuation width, i.e., 
\begin{equation}
W(L,t) = \langle {1 \over L^{d'}}\sum_x(h(x,t) - 
{\bar h(t)})^2 \rangle^{1/2}
\sim L^{\alpha}f(t/L^z),
\end{equation}
where $h(x,t)$ is the height of site $x$ on substrate at time 
$t$. $\bar h$, $L$, and $d'$ denote the mean height, system size, 
and substrate dimension, respectively. The angular brackets 
stand for statistical average. 
The scaling function behaves as $f(x) \rightarrow \hbox{const}$ 
for $x \gg 1$, and $f(x) \sim x^{\beta}$ for $x \ll 1$ with 
$z=\alpha/\beta$. The exponents $\alpha$, $\beta$ and $z$ are 
called the roughness, the growth, and the dynamic exponents, 
respectively. \\ 

Recently the problem of the pinning-depinning (PD) transition  
of interfaces in random media has also attracted interests in association 
with the problem of dynamics of fluctuating interfaces. 
Examples include the dynamics of domain boundaries of random Ising 
spin systems after being quenched below the critical temperature~\cite{rfield}, 
wetting immiscible displacement of one fluid by another 
in a porous medium~\cite{porous}, 
pinning flux lines in type-II superconductors~\cite{ertas}, 
fluid imbibition in paper~\cite{boston}, $etc$. 
In the problem of the PD transition, 
interface is pinned when external driving force $F$ 
is weaker than pinning strength induced by random media, 
while it moves with a certain velocity $v$ when the force $F$ 
is greater than the pinning strength. Thus there exists a 
threshold of external applying force $F_c$ across which 
the PD transition occurs. The role of the order parameter is played 
by the mean velocity, $v=\langle{\sum_x \partial h(x,t) 
/ \partial t} \rangle / L^{d'}$.  
Accordingly, the velocity is zero for $F < F_c$, and 
increases for $F > F_c$ as 
\be 
v \sim (F-F_c)^{\theta}, 
\ee
where the exponent $\theta$ is called the velocity exponent.\\ 

The continuum equation for the dynamics of interfaces in random 
media may be written simply as~\cite{levine} 
\be
{{\partial h(x,t)}\over {\partial t}}=\nu \nabla^2 h
+F+\eta(x,h),   
\ee 
where $h(x,t)$ is the height of the interface at position $x$ 
at time $t$. The first term on the right-hand side is from 
the smoothening effect of surface tension, the second 
term the uniform driving force, and the third a random force 
with short range correlations, satisfying $\langle \eta(x,h) 
\rangle =0$ and $\langle \eta(x, h) \eta(x',h') \rangle = 
2D \delta(x-x') \delta(h-h')$ with noise strength $D$. 
The above equation, called the quenched Edwards-Wilkinson (QEW) 
equation, would be relevant to the dynamics of the domain wall 
in random magnetic systems. More generally, recently a new continuum 
equation was introduced \cite{dhar}, which includes a nonlinear term 
${\lambda \over 2}(\nabla h)^2$ induced from the anisotropic 
property of the pinning strength. Thus the equation is replaced by 
\be
{{\partial h(x,t)}\over {\partial t}}=\nu \nabla^2 h
+{\lambda \over 2}(\nabla h)^2+F+\eta(x,h),   
\ee 
which is called the quenched Kardar Parisi Zhang (QKPZ) equation. 
The QKPZ equation leads to a different universality 
class from the QEW equation. Recently several stochastic 
models in the QKPZ universality class have been introduced
[9,11].  
From the models, it has been naturally concluded that the 
surface at the threshold of the PD transition 
$F_c$ can be described by the directed percolation 
(DP) cluster spanned perpendicularly to the surface 
growth direction in 1+1 dimensions. The roughness exponent 
$\alpha$ of the interface is given as the ratio of the correlation 
length exponents $\nu_{\perp}$ and $\nu_{\parallel}$ 
of the DP cluster in the transversal and the longitudinal 
direction that is $\alpha=\nu_{\perp}/\nu_{\parallel} \approx 
0.63$.\\ 

The dynamics of a single flux line in type-II superconductor 
with random impurities can also 
be understood in a similar manner used in the dynamics of 
surface growth. The main difference between them 
lies in that the flux line is an one-dimensional chain embedded 
in three dimensions rather than in two dimensions. 
Thus the roughness of the flux line is quantified in two different 
directions, the longitudinal and the transversal 
to the direction of its averaged velocity. 
Recently the continuum equations for the flux line dynamics in 
each directions were derived by Ertas and Kardar \cite{ertas}, 
which are coupled each other and look very complicated in 
general. They obtained the roughness and the dynamic exponents 
for various cases of the coupled equations; however, there 
still remain several cases that the roughness and dynamic 
exponents are not determined.  
In this paper, we will introduce a simple self-organized 
stochastic model, which we believe is relevant to the dynamics 
of the flux line. 
The numerical results we obtain from the stochastic model 
will compensate for the roughness and the dynamic exponents for 
the unknown cases. \\

The stochastic model we introduce in this paper is defined as follows. 
First, we consider a body centered cubic (bcc) lattice, 
in which an elastic string runs along the $x$-direction 
as shown in Fig.~1. The discrete version of the elastic 
string is composed of $L$-massless beads (black dots) which  
locate at nearest neighboring sites of one another 
and they are connected with strings. 
The configuration looking a zig-zag as shown in Fig.1 
is regarded as the initial flat configuration.
The total length of the string is equal to $\sqrt{3} aL/2$, 
where $a$ is a unit lattice constant of the bcc lattice 
and $L$ is the total number of beads in the system.  
The elastic string does not run backwardly to the $x$-direction, 
so that ($y,z$) positions of the bead for each $x$ are 
specified by single values. 
Each bead can be updated only in either positive 
$y$ or positive $z$-directions according to the following rule.   
First, two random numbers are assigned on each bead, 
one of which is for positive $y$-direction and the other is for 
positive $z$-direction. The random numbers for 
the $y$-direction are uniformly distributed between $[0,p]$ 
where $p \le 1$ and the ones for the $z$-direction are in $[0,1]$. 
Next, a minimum random number is selected among the $2L$ random 
numbers of the entire system, through which 
we determine the bead and the direction to move. 
Then the bead having the minimum random number is 
updated by shifting its location by the lattice 
constant $a$ along the direction chosen.  
Next, the avalanche process of updating may occur on 
neighboring beads when the separation between the nearest neighboring 
beads along the string becomes larger than $\sqrt{3} a /2$. 
In that case, the nearest neighboring bead is also 
shifted by a lattice constant along the direction already 
selected through the minimum random number.  
The avalanche rule is then applied successively to next neighboring 
beads to hold the conservation of the separation between nearest 
neighboring beads. The dynamic rule we used is similar to the 
Sneppen dynamic rule \cite{sneppen} but that the updating can occur 
in two different directions. Accordingly, we may say that 
our model is a vector Sneppen model. \\
\begin{figure}
\centerline{\epsfxsize=2cm \epsfbox{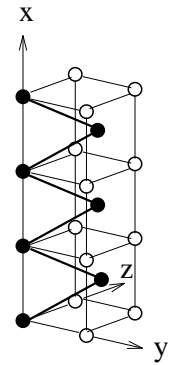}} 
\vspace{.5cm}
\caption{The initial flat configuration of the discrete 
version of elastic string. 
Each bead has two-component noises representing 
random pinning forces in $y$ and $z$ directions.} 
\label{fig1}
\end{figure}

The motion of the elastic string might be written by the 
following continuum equations, 
\begin{eqnarray}
&& {\partial_t y(x,t)}=\nu_y \partial_x^2 y
+{\lambda_y \over 2}(\partial_x y)^2+\eta_y(x,y,z), \nonumber \\
&& {\partial_t z(x,t)}=\nu_z \partial_x^2 z
+{\lambda_z \over 2}(\partial_x z)^2+\eta_z(x,y,z), 
\end{eqnarray}  
which are coupled via the quenched noise. 
The justification of the above equations associated with 
the stochastic model is based on that our model is similar to the 
Sneppen model but having the two different directions and the 
Sneppen model is a self-organized stochastic model belonging 
to the QKPZ universality in 1+1 dimensions. Even though 
the Sneppen model does not exhibit the PD transition, but 
the model gives the correct values of the roughness and 
the dynamic exponent in the QKPZ universality. 
Accordingly, we think that the self-organized stochastic model 
we introduced in this paper could give the correct values 
of the roughness and the dynamic exponents for the 
coupled QKPZ equations, Eq.(5). \\ 

When $p < 1$, the beads are more likely to advance in 
the $y$-direction, because random numbers are accumulated 
in the smaller interval $[0,p]$. 
The ratio of the averaged velocities $v_z / v_y$ of 
the $z$ and $y$ directions is equal to $p$.      
Thus we can define the angle $\psi=\tan^{-1}p$ as 
the angle between the $y$-axis and the averaged propagating 
direction. Transforming the coordinate system $(y,z)$ by the angle 
$\psi$ into the new system $(r_{\parallel}, r_{\perp})$, 
the above equations can be rewritten in terms of 
$r_{\parallel}$ and $r_{\perp}$,  
where $r_{\parallel}$ is along the direction of the averaged velocity 
and $r_{\perp}$ is perpendicular to $r_{\parallel}$.  
The continuum equations in the $(r_{\parallel}, r_{\perp})$ coordinate 
system are written as  
\begin{eqnarray} 
{\partial_t r_{\parallel}(x,t)}&& =
\nu_{11}\partial_x^2 r_{\parallel}+\nu_{12}\partial_x^2 r_{\perp}+
{\lambda_{11}\over 2}(\partial_x r_{\parallel})^2+
\nonumber \\
&& {\lambda_{12}\over 2}(\partial_x r_{\perp})^2 
+{\lambda_{13}\over 2}(\partial_x r_{\parallel})(\partial_x r_{\perp})
+\eta_{\parallel}(x,r_{\parallel},r_{\perp}), \nonumber \\
{\partial_t r_{\perp}(x,t)}&& =
\nu_{21}\partial_x^2 r_{\parallel}+\nu_{22}\partial_x^2 r_{\perp}+
{\lambda_{21}\over 2}(\partial_x r_{\perp})^2+
\\
&& {\lambda_{22}\over 2}(\partial_x r_{\parallel})^2 
+{\lambda_{23}\over 2}(\partial_x r_{\perp})(\partial_x r_{\parallel})
+\eta_{\perp}(x,r_{\parallel},r_{\perp}), \nonumber  
\end{eqnarray}  
where $\nu_{ab}$ with $a,b \in 1$ or $2$ are functions of 
$\nu_y, \nu_z, \cos\psi$, and $\sin\psi$,
and $\lambda_{ab}$ with $a, b \in 1, 2$, or $3$ are of 
$\lambda_y, \lambda_z, \cos \psi$ and $\sin\psi$. 
$\eta_{\parallel}$ and $\eta_{\perp}$ are the random noises 
in the longitudinal and the transversal direction, respectively. 
When $p=1$, the advance of the elastic string in $y$ and $z$ directions 
are identical, and the above equations are reduced simply as 
\begin{eqnarray} 
{\partial_t r_{\parallel}(x,t)}&=&
\nu_{11}\partial_x^2 r_{\parallel}+
{\lambda_{11}\over 2}(\partial_x r_{\parallel})^2+
{\lambda_{12}\over 2}(\partial_x r_{\perp})^2 
+\eta_{\parallel}, \\  
{\partial_t r_{\perp}(x,t)}&=&
\nu_{22}\partial_x^2 r_{\perp}+
{\lambda_{23}\over 2}(\partial_x r_{\perp})(\partial_x r_{\parallel})
+\eta_{\perp}. 
\end{eqnarray}  
Note that $\lambda_{11}\ne 0$ in Eq.~(7) and 
Eq.(8) is invariant under $r_{\perp} \rightarrow -r_{\perp}$.
In order to obtain the roughness exponents for each direction, 
we consider the spatial correlation functions $C_{\parallel}$ 
and $C_{\perp}$ after saturation, 
\begin{eqnarray} 
C_{\parallel}(x,t) &=& 
\langle {1\over L}\sum_{x_1} \big(r_{\parallel}
(x+x_1,t)-r_{\parallel}(x_1,t)\big)^2 
\rangle^{1/2}, \nonumber \\ 
C_{\perp}(x,t) &=& \langle {1\over L}\sum_{x_1}
\big(r_{\perp}(x+x_1,t)-r_{\perp}(x_1,t)\big)^2 
\rangle^{1/2}, 
\end{eqnarray}
which behave as $C_{\parallel}(x) \sim x^{\alpha_{\parallel}}$ 
and $C_{\perp}(x) \sim x^{\alpha_{\perp}}$.
Next, in order to obtain the growth exponents, we  
consider the temporal correlation functions 
$\tilde C_{\parallel}$ and $\tilde C_{\perp}$, 
\begin{eqnarray} 
\tilde C_{\parallel}(t_2-t_1) & = & \langle 
{1\over L}\sum_x(r_{\parallel}(x,t_2)-\bar r_{\parallel}(t_2)
-r_{\parallel}(x,t_1)+\nonumber \\ 
&&\bar r_{\parallel}(t_1))^2 \rangle^{1/2},
\nonumber \\
\tilde C_{\perp}(t_2-t_1)&=& \langle 
{1 \over L}\sum_x(r_{\perp}(x,t_2)-\bar r_{\perp}(t_2)
-r_{\perp}(x,t_1)+\nonumber \\ 
&& \bar r_{\perp}(t_1))^2 \rangle^{1/2}, 
\end{eqnarray} 
where $t_1$ is taken as a saturation time. 
The correlation functions behave as 
$\tilde C_{\parallel} \sim (t_2-t_1)^{\beta_\parallel}$
and 
$\tilde C_{\perp} \sim (t_2-t_1)^{\beta_\perp}$
Numerical simulations are performed for the cases 
of $p=1$ and $p=1/2$. \\
 
For $p=1$, where Eqs.(7-8) are hold, the roughness 
exponents are measured as $\alpha_{\parallel} \approx 0.63$ 
and $\alpha_{\perp} \approx 0.50$, and the data 
are shown in Fig.2. 
The growth exponents are measured as $\beta_{\parallel}\approx 
0.64$ and $\beta_{\perp}\approx 0.53$, 
and the data are shown in Fig.3.   
From the numerical results, the dynamic exponents are 
obtained as $z_{\parallel}\approx 0.99$ and 
$z_{\perp}\approx 0.95$, which suggest that  
the true dynamic exponents be $z_{\parallel}=z_{\perp}=1$. 
This result is attributed to that the coherent effect propagates 
along the string and the chemical distance between two points 
on string remains invariant through the process of the 
dynamics \cite{havlin}. 
\begin{figure}
\centerline{\epsfxsize=8cm \epsfbox{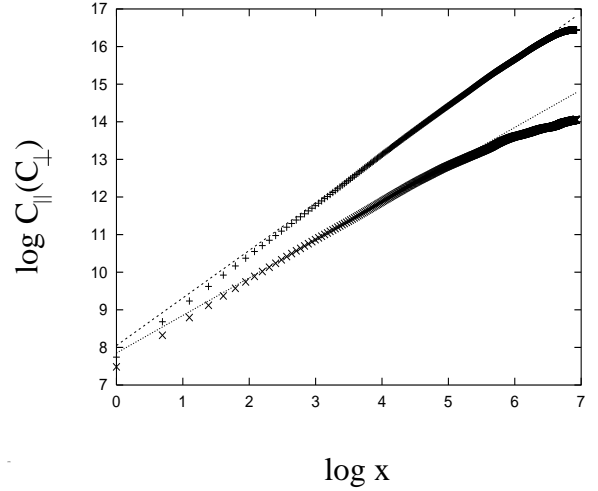}} 
\vspace{.5cm}
\caption{
Plot of $C_{\parallel}$ ($C_{\perp}$) versus $x$ 
for $p=1$ in double logarithmic scales. 
The simulations are performed for the system size $L=2048$. 
The lines obtained from the least square fits have 
the slopes $1.26$ (top) and $1.00$} 
\label{fig2}
\end{figure}
\begin{figure}
\centerline{\epsfxsize=8cm \epsfbox{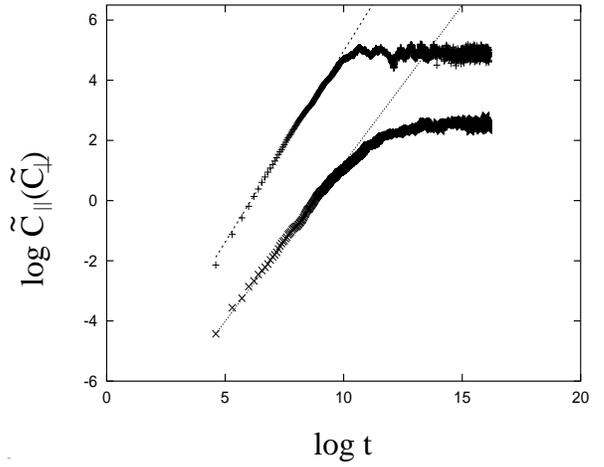}} 
\vspace{.5cm}
\caption{
Plot of $\tilde C_{\parallel}$ ($\tilde C_{\perp}$) 
versus time for $p=1$ in double logarithmic scales. 
The simulations are performed for the system size $L=2048$. 
The lines obtained from the least square fit 
have the slopes $1.27$ (top) and $1.05$} 
\label{fig3}
\end{figure}

Next for $p=1/2$, the two spatial correlation functions 
for each direction are less distinctive than 
the case of $p=1$ as shown in Fig.4. 
The roughness exponents are obtained as 
$\alpha_{\parallel} \approx 0.60$ and 
$\alpha_{\perp} \approx 0.54$. 
For the growth exponents, the time correlation function 
$\tilde C_{\parallel}$ in the longitudinal direction exhibits 
the power-law dependent behavior against time 
with the exponent $\beta_{\parallel}\approx 
0.64$. However, for the transversal direction, the data 
do not exhibit a simple power-law type behavior, 
rather they show a crossover behavior from 
$\beta_{\perp}\approx 0.50$ to $\beta_{\perp} < 0.50$ in Fig. 5.   
From the numerical results, it is likely that 
the roughness exponents are $\alpha_{\parallel} 
=0.63$ and $\alpha_{\perp}=0.5$, and 
the dynamic exponents are $z_{\parallel}=z_{\perp}=1$ 
even for $0 < p < 1$. Note that when $p=0$, the dynamics 
in the transversal direction does not exist at all 
in our model. Accordingly, the numerical results 
indicate that the roughness and dynamic exponents 
do not depend on the angle $\psi$. 
Also the numerical values of the roughness and dynamic exponents 
suggest that the dynamics in the longitudinal direction belongs to the 
directed percolation depinning universality class (the QKPZ 
universality class) in 1+1 directions.  
On the other hand, for the transversal direction, 
the numerical values $\alpha_{\perp}
\approx 0.5$ and $z_{\perp}\approx 1$ coincide with 
the ones in the anisotropic QKPZ universality class [10,14], 
even though the dynamics equation, Eq.(8), does not include
the symmetric breaking term, $(\partial_x r_{\perp})^2$. 
We think that this coincidence is due to the presence 
of the term proportional to $\partial_x r_{\perp}$ 
in Eq.(8).\\  
\begin{figure}
\centerline{\epsfxsize=8cm \epsfbox{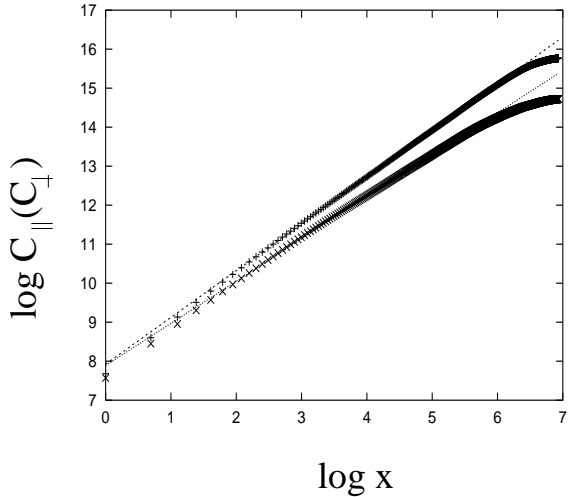}} 
\vspace{.5cm}
\caption{Plot of $C_{\parallel}$ ($C_{\perp}$) versus $x$ 
for $p=1/2$ in double logarithmic scales. 
The simulations are performed for the system size $L=2048$. 
The lines obtained from the least square fit have 
the slopes $1.20$ (top) and $1.08$.} 
\label{fig4}
\end{figure}
\begin{figure}
\centerline{\epsfxsize=8cm \epsfbox{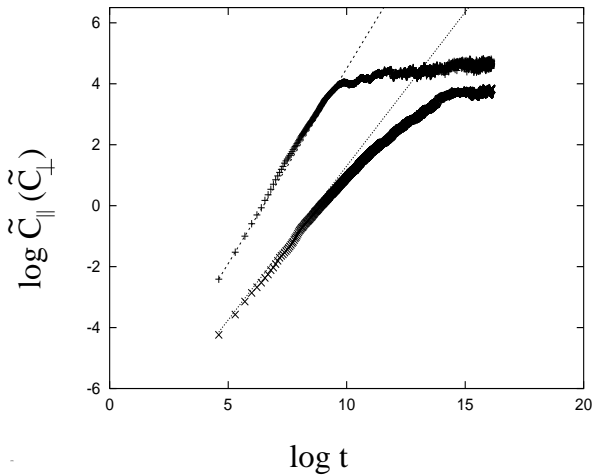}} 
\vspace{.5cm}
\caption{
Plot of $\tilde C_{\parallel}$ ($\tilde C_{\perp}$) 
versus time for $p=1/2$ in double logarithmic scales. 
The simulations are performed for the system size $L=2048$. 
The lines have the slopes $1.28$ (top) and $1.00$} 
\label{fig5}
\end{figure}
 
In summary, we have introduced a stochastic model associated with 
the dynamics of the flux line in quenched media at the depinning threshold.  
The conservation of the total length of string through the dynamics
in the current stochastic model leads to $z=1$ 
in both the longitudinal and the transversal direction. 
The roughness exponents were obtained 
as $\alpha_{\parallel} \approx 0.63$ and $\alpha_{\perp}\approx 0.5$, 
respectively. \\

This work was supported in part by the Non-Directional Research 
Fund, Korea Research Foundation, in part by the KOSEF through the 
SRC program of SNU-CTP, and in part by the Ministry of Education 
(BSRI 97-2409), Korea. \\

\end{multicols}
\end{document}